\documentclass[10pt,a4paper]{article}
\usepackage{amsmath}
\usepackage{amsfonts}
\usepackage{amssymb}
\usepackage{graphicx}
\usepackage{textcomp}
\usepackage{multicol}
\usepackage{url}
\usepackage{lipsum}
\usepackage{setspace}
\usepackage{comment}
\usepackage{color}
\usepackage[numbers,sort&compress]{natbib}
\usepackage[usenames,dvipsnames,svgnames,table]{xcolor}
\usepackage[lofdepth,lotdepth,caption=false]{subfig}
\usepackage[pdfstartview=FitH]{hyperref}
\usepackage{hyperref}
\newsavebox{\bigleftbox}
\hypersetup{
 colorlinks,
 citecolor=Blue,
 linkcolor=Blue,
 urlcolor=Blue
}

\makeatletter
 \def\footnoterule{\kern-3\p@
   \noindent\hrulefill \kern 2.8\p@} 
\makeatother
\usepackage[left=3.00cm, right=3.00cm, top=2.00cm, bottom=2.00cm]{geometry}

\title{\textbf{Structural, Electronic, and Li-ion Adsorption Properties of PolyPyGY Explored by First-Principles and Machine Learning Simulations: A New Multi-Ringed 2D Carbon Allotrope}}
\author{
   K. A. L. Lima$^{1,2}$, 
   D. A. da Silva$^{3,4}$, 
   G. D. Amvame Nze$^{3,4}$,  \\
   F. L. Lopes de Mendonça$^{3,4}$,
   M. L. Pereira, Jr$^{3}$, and
   L. A. Ribeiro, Jr$^{1,2,\S}$
	}
\date{}

\begin{document}
    \maketitle
	\vspace{-0.6cm}
	\begin{center}\small
     \textit{$^{1}$Institute of Physics, University of Bras\'ilia, 70910-900, Bras\'ilia, Brazil}\\
     \textit{$^{2}$Computational Materials Laboratory, LCCMat, Institute of Physics, University of Bras\'ilia, 70910-900, Bras\'ilia, Brazil}\\
     \textit{$^{3}$University of Bras\'ilia, College of Technology, Department of Electrical Engineering, Bras\'ilia, Brazil}\\
     \textit{$^{4}$Professional Postgraduate Program in Electrical Engineering (PPEE), Department of Electrical Engineering, College of Technology, University of Bras\'{i}lia, Bras\'{i}lia, Federal District, Brazil} \\
    \phantom{.}\\ \hfill
        $^{\S}$\url{ribeirojr@unb.br}\hfill
		\phantom{.}
	\end{center}
	

\onehalfspace

\noindent\textbf{Abstract: Two-dimensional (2D) carbon materials have been intensively investigated because of their distinctive structural framework and electronic behaviors as alternatives in energy conversion and storage applications. This study proposes a novel 2D carbon allotrope, Polymerized Pyracyclene Graphyne (PolyPyGY), characterized by a multi-ringed structure with 4-, 5-, 6-, 8-, and 16-membered rings comprising a porous structure. Using first-principles calculations and machine-learning techniques, we explore its structural, electronic, mechanical, optical, and lithium-ion storing properties. The vibrational properties assessed through the density functional perturbation theory framework confirm its structural stability. Moreover, ab initio molecular dynamics simulations at 1000 K demonstrate its thermal resilience, with no bond breaking or reconfiguration observed. The electronic band structure reveals a metallic nature, and the material exhibits anisotropic elastic properties, with Young's modulus varying between 421 and 664 GPa, suggesting good mechanical stability. Furthermore, lithium diffusion studies indicate low energy barriers (0.05-0.9 eV) and a high diffusion coefficient ( $>$ 6 $\times$ 10$^{-6}$ cm$^{2}$/s), along with a stable open circuit voltage of 1.2 V. These results highlight PolyPyGY's potential as a highly effective and durable anode material for lithium-ion batteries, featuring rapid Li-ion diffusion, stable intercalation, and consistent performance during charge and discharge cycles.}

\section{Introduction}

Two-dimensional (2D) carbon-based materials have been extensively studied attributed to their exceptional structural and optoelectronic characteristics \cite{hirsch2010era,nasir2018carbon,lu2013two,tiwari2016magical}, including high surface area \cite{yi2018microporosity}, excellent electrical \cite{fan2017new} and thermal conductivity \cite{pereira2021investigating,yue2017thermal}, and good mechanical resilience \cite{zhao2013mechanical}. These properties make them promising candidates for various advanced technological applications \cite{singhal2022carbon}. Among these applications, using such materials as active layers in lithium-ion batteries (LIBs) stands out \cite{rajkamal2019carbon,lherbier2018lithiation,wang2021carbon,ali2023advancements}, where lithium storage capacity and ion diffusion speed are crucial factors for optimizing battery efficiency and durability.

The porous structure of 2D carbon materials, composed of varying-sized rings, enables a higher lithium adsorption capacity while facilitating Li diffusion across the material's surface \cite{cocco2013three}. These porous materials include allotropes such as graphyne \cite{desyatkin2022scalable},  graphdiyne \cite{gao2019graphdiyne}, and graphenylene \cite{song2013graphenylene}, which, by presenting various carbon ring configurations, provide a structural network capable of stabilizing lithium insertion without compromising the material's mechanical integrity \cite{yu2013graphenylene}. This feature is relevant for enhancing the capacity of LIBs, which are essential components in several energy storage technologies \cite{ruby2024recent,li201830,wakihara2001recent,kim2019lithium,scrosati2011lithium,blomgren2016development,manthiram2017outlook}.

Recent studies on new 2D carbon allotropes have emphasized the need for structures that combine high adsorption capacity with optimized lithium diffusion barriers \cite{gomez2024tpdh,wang2018popgraphene,santos2024proposing,li2021two,liu2024novel,li2017psi,cai2023lc,lu2022new,cai2021net,cheng2021two,you20242d,liu2017new,wang2022thgraphene,santos2024photh}. Such characteristics can be achieved through structural configurations that maximize porosity, offering diverse adsorption sites and enabling efficient diffusion pathways. The nanomaterial proposed in this study, Polymerized Pyracyclene Graphyne (PolyPyGY), is a novel multi-ringed 2D all-carbon structure featuring 4, 5, 6, 8, and 16-membered rings, emerging as a promising response to these demands. Its porous structure with good thermal and mechanical stability has the potential to overcome limitations observed in conventional graphene-based allotropes when designing lithium-ion battery anodes \cite{liu2022applications}.

Herein, we investigate the structural, electronic, mechanical, optical, and lithium-ion adsorption properties of PolyPyGY using first-principles calculations and machine learning (ML) techniques. We examine its stability through phonon calculations and ab initio molecular dynamics (AIMD) simulations, characterize its electronic band structure, and assess its lithium diffusion barrier, adsorption capacity, and open-circuit voltage. This newly proposed 2D carbon allotrope originates from assembling dehydrogenated pyracyclene molecules \cite{lo1968pyracyclene,trost1967pyracylene}. These molecules are interconnected by carbon atoms forming triple bonds in the $x$-direction, which create a highly stable and unique porous framework. This particular choice for including the triple bonds enables the formation of 16-membered rings with an adequate porous size for Li-ion storage. The design incorporates an innovative 8-membered ring conjunction that bridges neighboring pyracyclene-based motifs, resulting in a distinct topology. This assembly leads to a porous and robust architecture that balances mechanical stability with ample space for lithium adsorption. The combination of pyracyclene motifs and graphyne-like linkages enhances the material's electronic and adsorption properties, making PolyPyGY a promising high-performance anode material for LIBs, offering an ideal combination of fast Li-ion diffusion, stability, and storing capacity.

\section{Methodology}

To investigate the underlying physical, chemical, and lithium-ion adsorption properties of PolyPyGY, we employed first-principles calculations based on density functional theory (DFT) using the CASTEP code \cite{clark2005first}. The initial atomic structure of PolyPyGY was optimized using the Perdew-Burke-Ernzerhof (PBE) functional within the generalized gradient approximation (GGA) scheme for the exchange-correlation functional \cite{perdew1996generalized}. A plane-wave energy cutoff of 450 eV and a Monkhorst-Pack k-point mesh of $10\times 10\times 1$ were applied for Brillouin zone sampling. Convergence criteria for energy and forces are $1.0 \times 10^{-5}$ eV and $1.0 \times 10^{-3}$ eV/\r{A}, respectively. Structural relaxation employed periodic boundary conditions, ensuring minimal residual forces and pressure below 0.01 GPa. The vacuum space was set to 30 \r{A} in the out-of-plane direction to avoid layer interaction. Phonon calculations were performed to confirm the structural stability of PolyPyGY, using density functional perturbation theory (DFPT), ensuring the absence of imaginary frequencies across the phonon dispersion spectrum.

Thermal stability was assessed via AIMD simulations at 1000 K, using a simulation time of 5 ps and a timestep of 1 fs. The AIMD simulations were conducted in the NVT ensemble with a Nosé-Hoover thermostat \cite{nose1984unified}. Throughout the simulation, we monitored the structural integrity to detect any bond breaking or atomic reconfiguration that could indicate thermal instability. The electronic structure of PolyPyGY was analyzed by calculating the electronic band (using a $k$-point grid of $10\times10\times1$) structure and density of states (DOS, using a $k$-point grid of $20\times20\times1$). The partial density of states (PDOS) was also projected to identify contributions from specific atomic orbitals, providing insights into the material's conductive properties. Optical properties were explored under an external electric field using complex dielectric constant calculations \cite{lima2023dft}. 

Lithium adsorption studies were conducted by placing Li atoms at various adsorption sites on the PolyPyGY surface, followed by geometry optimization to identify energetically favorable configurations. The adsorption energy (E$_{\text{ads}}$) was calculated using the formula 
$E_{\text{ads}} = E_{\text{Li+PolyPyGY}} - \left( E_{\text{PolyPyGY}} + E_{\text{Li}} \right)$, where $E_{\text{Li+PolyPyGY}}$ is the total energy of PolyPyGY with adsorbed Li, $ E_{\text{PolyPyGY}}$ is the energy of pristine PolyPyGY, and $E_{\text{Li}}$ is the energy of an isolated Li atom. We computed the diffusion barrier along possible migration paths using the nudged elastic band (NEB) method within CASTEP to investigate lithium diffusion \cite{makri2019preconditioning,barzilai1988two,bitzek2006structural}. The diffusion coefficient was estimated from these results, providing insights into lithium mobility on the PolyPyGY surface.

The mechanical behavior of PolyPyGY was investigated through classical reactive molecular dynamics simulations using the Large-scale Atomic/Molecular Massively Parallel Simulator (LAMMPS) software \cite{plimpton1995fast}. A reactive force field tailored specifically for PolyPyGY was derived from a moment tensor potential (MTP) \cite{mortazavi2021first}, which was fitted to explore its mechanical properties with the assistance of the Machine Learning Interatomic Potential (MLIP) package \cite{novikov2020mlip,podryabinkin2017active}. AIMD simulations were conducted following the same parameters used previously, employing the Vienna Ab initio Simulation Package (VASP) \cite{kresse1993ab,kresse1996efficiency,kresse1994norm}. The training dataset for the MTP-based MLIP was constructed using stress-free and uniaxially strained supercells at various temperatures, an approach that has been successfully applied in recent studies to generate MLIPs for complex materials \cite{mortazavi2023machine,mortazavi2025exploring,mortazavi2023atomistic,mortazavi2020efficient,mortazavi2023electronic,mortazavi2024goldene,lima2024th,mortazavi2022machine,mortazavi2022anisotropic}.

For the MTP, a cutoff distance of 3.5 \r{A} was established, and training was performed using a two-step passive training approach, as described in recent literature \cite{mortazavi2024goldene}. Newton’s equations of motion were integrated using the velocity Verlet algorithm with a timestep of 0.05 fs. Thermalization and stress relaxation were achieved through the thermostat and NPT ensemble simulations throughout 100 ps. The elastic properties and fracture behavior of the PolyPyGY sheet were assessed at 300 K, where the material was subjected to uniaxial tension using a constant engineering tensile strain rate of 10$^{-6}$ fs$^{-1}$.

\section{Results}

\begin{figure}[!htb]
    \centering
    \includegraphics[width=0.9\linewidth]{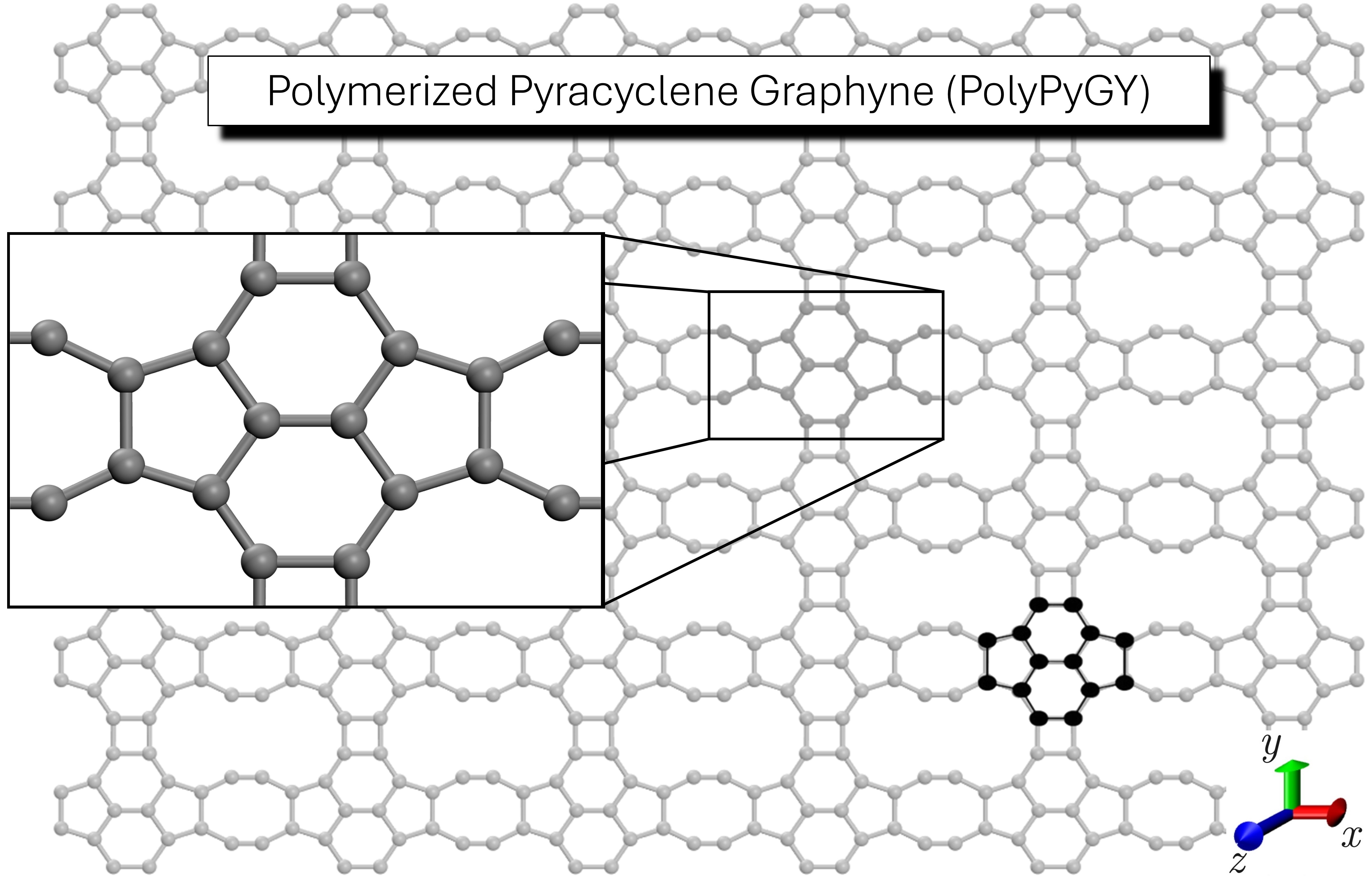}
    \caption{Atomic structure of PolyPyGY. The unit cell, outlined in black, has dimensions of $a=9.46$ \r{A} ($x$-direction) e $b=6.08$ \r{A} ($y$-direction). The black spheres denote the structure of a dehydrogenated pyracyclene molecule.}
    \label{fig:system}
\end{figure}

The atomic structure of PolyPyGY is presented in Figure \ref{fig:system}. As depicted, the lattice is composed of 4-, 5-, 6-, 8-, and 16-membered rings of carbon atoms, forming a highly porous structure. The unit cell, highlighted in black, contains 18 atoms with dimensions $a=9.46$ \r{A} ($x$-direction) e $b=6.08$ \r{A} ($y$-direction). The C-C bond lengths vary slightly across the structure, with typical values ranging between 1.38 \r{A} to 1.48 \r{A}, corresponding to different bonding environments between carbon atoms in the various ring sizes. Crystallographically, PolyPyGY belongs to the PMMM space group space group and adopts a D2H-1 symmetry.

The calculated formation energy for PolyPyGY is $-8.44$ eV/atom, indicating good energetic stability. This value is comparable to other 2D carbon allotropes, such as biphenylene network ($-7.42$ eV/atom) \cite{luo2021first}, graphene ($-8.83$ eV/atom) \cite{skowron2015energetics}, sun-graphyne ($-8.57$ eV/atom) \cite{tromer2023mechanical}, irida-graphene (-6.96 eV/atom) \cite{junior2023irida}, and dodecanophene ($-8.19$ eV/atom) \cite{lima2024dodecanophene}. It is significantly more stable than other porous allotropes, like graphdiyne ($-0.77$ eV/atom) and $\gamma$-graphyne ($-0.92$ eV/atom) \cite{zhao2013two}. The lattice of PolyPyGY appears perfectly planar, with no observed buckling deformations. PolyPyGY's multi-ringed architecture, especially larger 16-membered rings, significantly contributes to its high porosity and large surface area, which are advantageous for lithium-ion storage and diffusion applications. This unique atomic arrangement also supports robust mechanical stress, as discussed later.

\begin{figure}[!htb]
    \centering
    \includegraphics[width=0.8\linewidth]{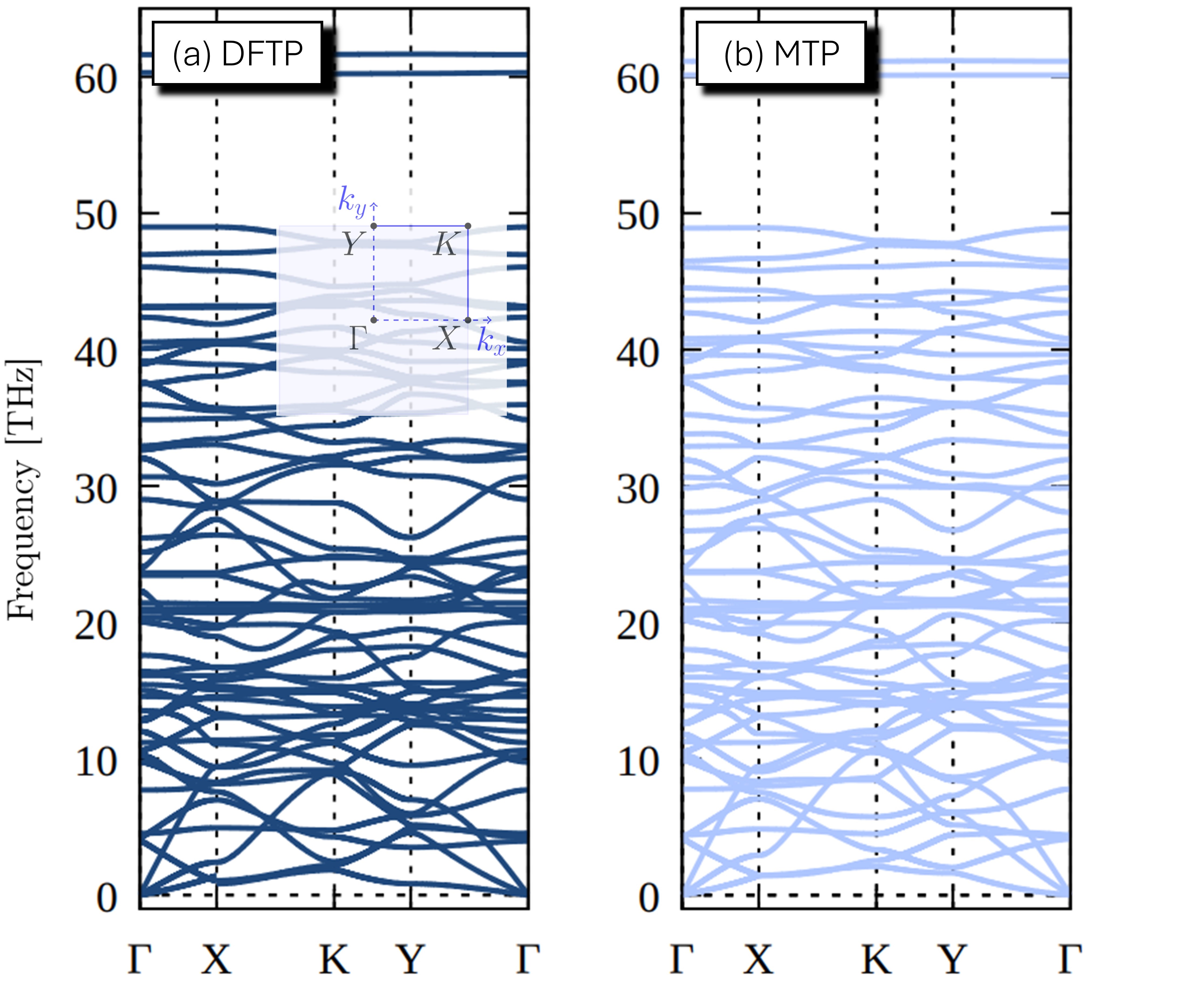}
    \caption{Phonon dispersion analysis of PolyPyGY calculated using (a) DFPT and (b) MTP methods. The inset highlights the first Brillouin zone.}
    \label{fig:phonons}
\end{figure}

Figure \ref{fig:phonons} presents the phonon dispersion results for PolyPyGY, showcasing its dynamic stability. Figure \ref{fig:phonons}(a) shows the phonon calculations performed using DFPT. In contrast, Figure \ref{fig:phonons}(b) displays results from phonon calculations based on the MTP method \cite{mortazavi2020exploring}. In both panels, the absence of imaginary frequencies across the phonon spectrum suggests the dynamic stability of PolyPyGY. The MTP-based phonon calculations in Figure \ref{fig:phonons}(b) strongly agree with the DFPT results in Figure \ref{fig:phonons}(a), validating the accuracy of the generated MLIP for PolyPyGY. The alignment of results from DFPT and MTP further supports the use of ML-based potentials here and in future studies focused on the mechanical and thermal behavior of PolyPyGY using classical reactive MD simulations. Also, the consistency between these two methods can enable efficient simulations of larger systems and longer timescales without sacrificing accuracy.

\begin{figure}[!htb]
    \centering
    \includegraphics[width=0.8\linewidth]{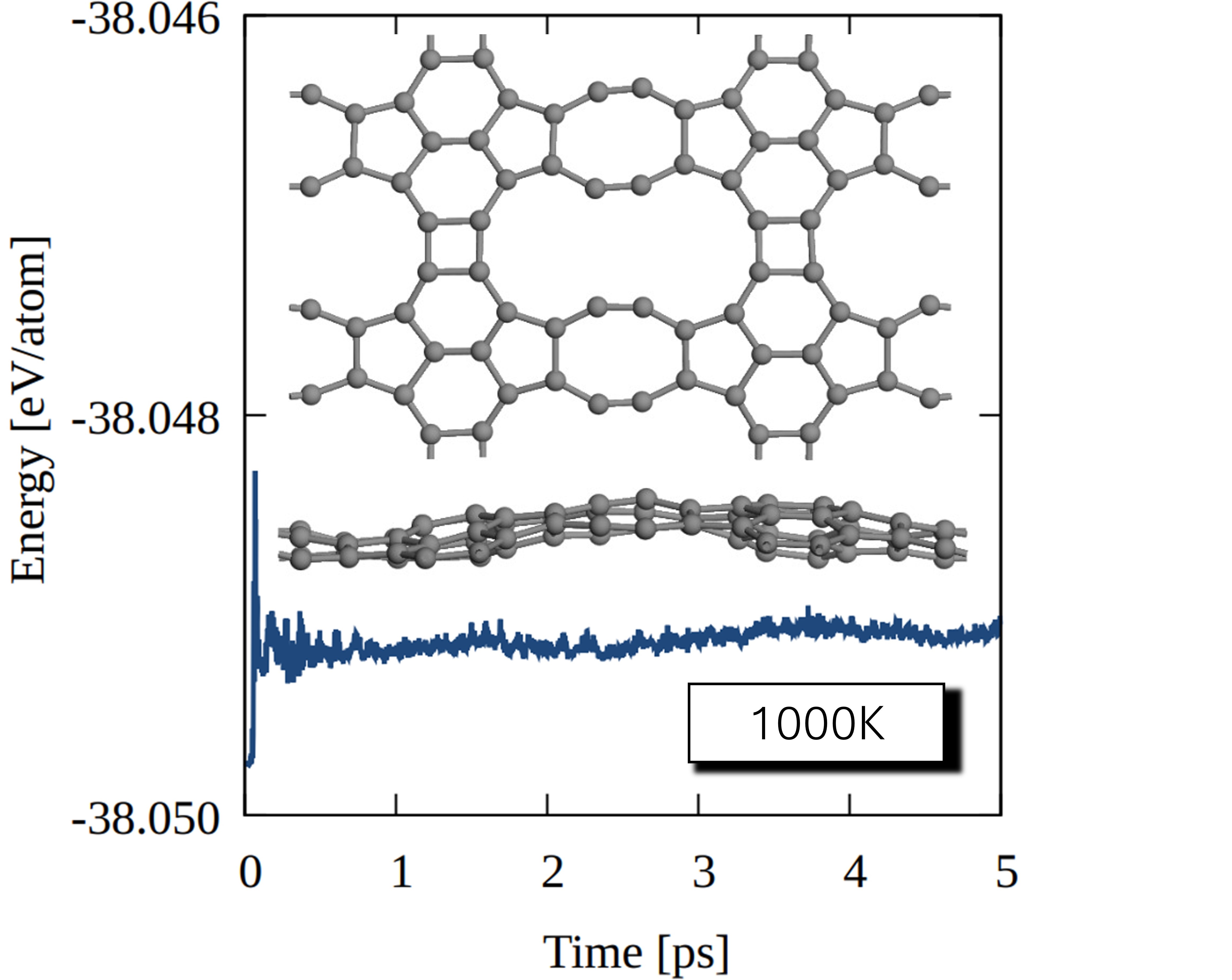}
    \caption{The AIMD simulation results at 1000 K, depicting the lattice total energy as a function of time. Insets show top and side views of the structure for the final AIMD snapshot at 5 ps.}
    \label{fig:aimd}
\end{figure}

In Figure \ref{fig:aimd}, the thermal stability of PolyPyGY is assessed through AIMD simulations at 1000 K. Insets show top and side views of the structure for the final AIMD snapshot at 5 ps. The absence of bond breaking or structural reconfiguration in the insets indicates that PolyPyGY remains thermally stable under high-temperature conditions. This stability is critical for practical applications, especially in environments where materials are subjected to significant thermal stress, such as LIB anodes. The insets also show that the lattice maintains its integrity without any observable in-plane and drastic out-of-plane deformation.

\begin{figure*}[!htb]
    \centering
    \includegraphics[width=\linewidth]{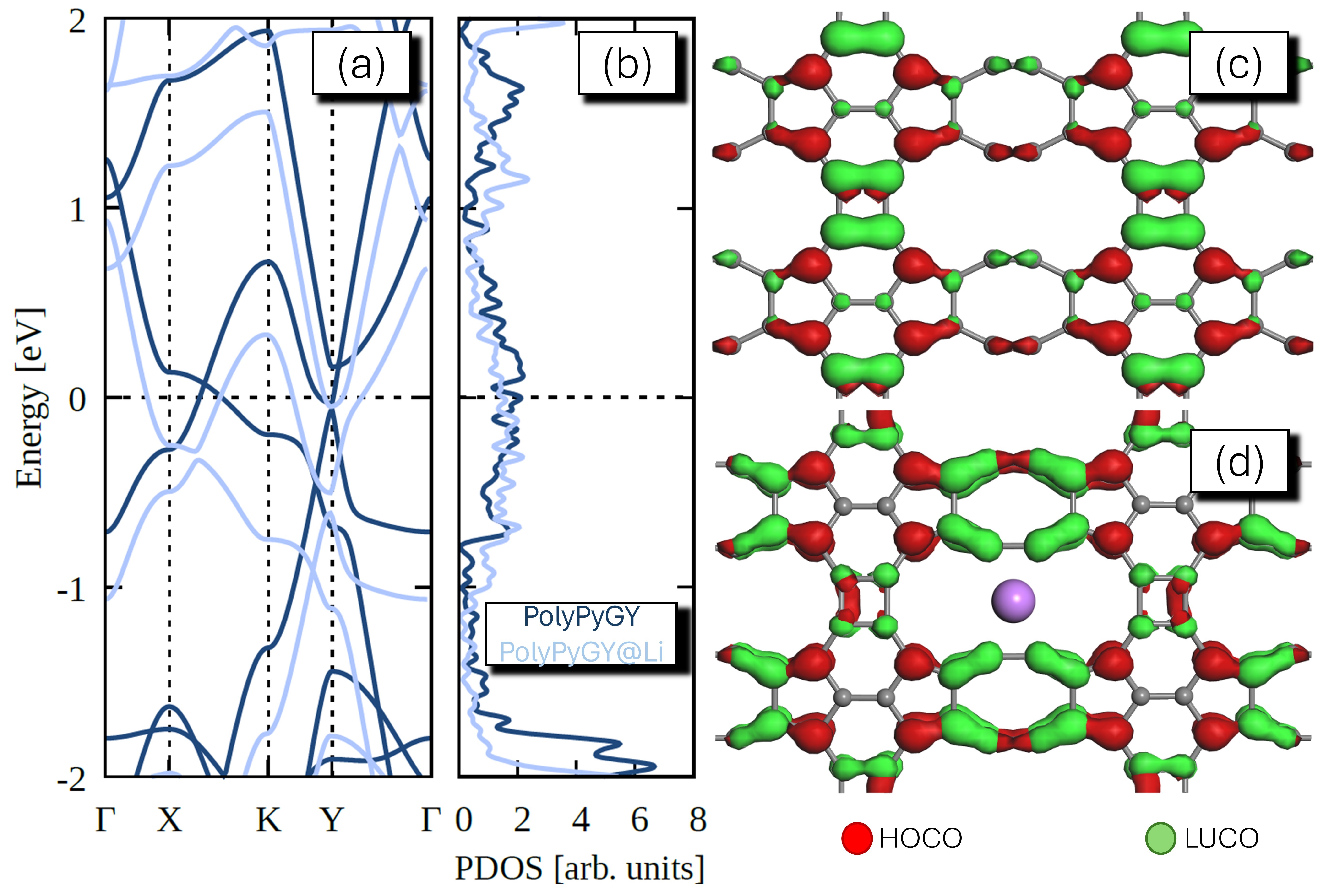}
    \caption{Electronic structure of PolyPyGY. (a) Band structure along high-symmetry $k$-paths and (b) PDOS. Panels (c) and (d) depict the Highest Occupied Crystalline Orbital (HOCO, red) and Lowest Unoccupied Crystalline Orbital (LUCO, green) distributions for PolyPyGY and PolyPyGY@Li, respectively.}
    \label{fig:bands-orbitals}
\end{figure*}

Now, we turn to the electronic properties of the proposed nanomaterial. Figure \ref{fig:bands-orbitals} shows these properties for pure PolyPyGY (dark blue) and the lattice containing one adsorbed Li atom (PolyPyGY@Li, light blue). In Figure \ref{fig:bands-orbitals}(a), the electronic band structure is shown along high-symmetry $k$-paths, revealing that PolyPyGY exhibits a metallic nature. PolyPyGY can conduct electrons efficiently, making it suitable for applications requiring controlled electronic transport. Figure \ref{fig:bands-orbitals}(b) presents the PDOS, where only contributions from $p$ atomic orbitals are visible for the depicted range of energies. The PDOS confirms the dominant orbital contributions, such as carbon p-orbitals near the Fermi level, consistent with the PolyPyGY's metallic character. 

The band structure of PolyPyGY exhibits a downward shift in energy levels upon lithium adsorption. This shift arises from the charge transfer process, where lithium donates electrons to the PolyPyGY lattice. As a result, the Fermi level value increases, stabilizing the electronic structure by redistributing charge within the conduction band. This behavior confirms the material's ability to accommodate lithium ions and underscores its suitability as a potential anode material for lithium-ion batteries. 

Figures \ref{fig:bands-orbitals}(c,d) show the highest occupied crystalline orbital (HOCO, red) and the lowest unoccupied crystalline orbital (LUCO, green). The spatial distribution of these orbitals indicates that HOCO delocalizes on the 5- and 8-membered rings while LUCO concentrates in 4- and 6-membered rings, revealing distinct electron and hole transport channels across the material. The HOCO and LUCO distributions suggest that PolyPyGY could support efficient and controllable charge transport, further supporting its potential use in conductive layers for energy storage devices. The orbital distribution in pristine PolyPyGY (Figure \ref{fig:bands-orbitals}(a)) undergoes significant changes upon lithium adsorption, as shown in Figure \ref{fig:bands-orbitals}(d). the polarization effects induced by the lithium atom and minor structural distortions in the material further enhance the confinement of electronic states in its vicinity. Consequently, the adsorption process increases the localization of the LUCO around the lithium atom.  

\begin{figure*}[!htb]
    \centering
    \includegraphics[width=0.8\linewidth]{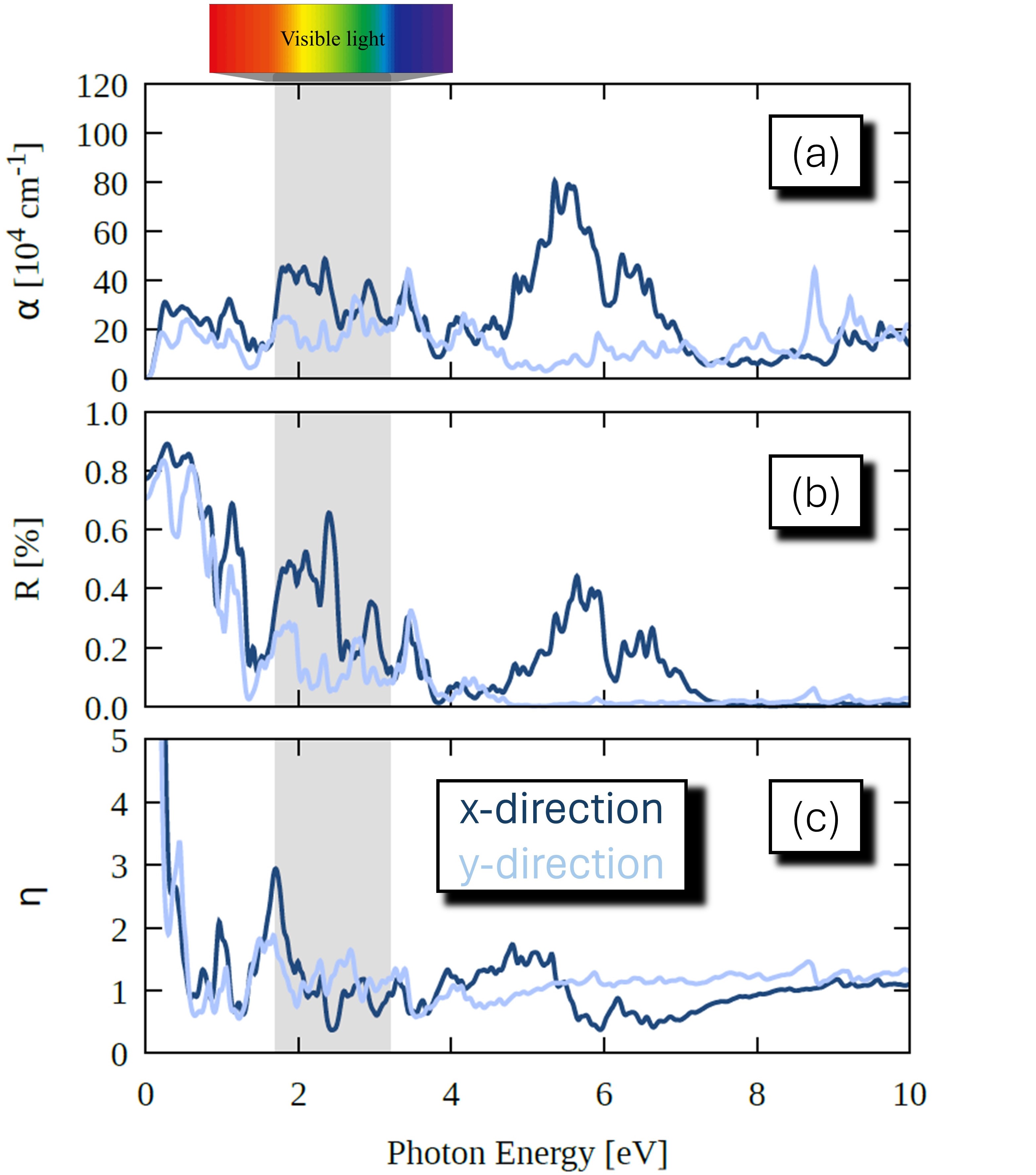}
    \caption{Optical properties of PolyPyGY. (a) Absorption coefficient ($\alpha$), (b) reflectivity ($R$), and (c) refractive index ($\eta$) as a function of photon energy and for light polarized along the $x$-direction (dark blue) and $y$-direction (light blue).}
    \label{fig:optical}
\end{figure*}

PolyPyGY exhibits interesting optical characteristics, with polarization-dependent profiles along the $x$-direction (dark blue) and $y$-direction (light blue), as illustrated in Figure \ref{fig:optical}. In Figure \ref{fig:optical}(a), the absorption coefficient ($\alpha$) is plotted against photon energy, revealing broad optical responses with peaks in the visible to ultraviolet (UV) range. The absorption coefficient reaches values exceeding 10$^{4}$ cm$^{-1}$, indicating the material's high absorption efficiency at these wavelengths. These intense and broad absorption trends suggest that PolyPyGY could be advantageous in optoelectronic applications, where effective light capture in a broad spectrum range is desirable.

Figure \ref{fig:optical}(b) shows reflectivity ($R$) results. The material exhibits low reflectivity across the studied energy range, remaining below 0.9, which implies that a significant amount of incident light is transmitted rather than reflected. The clear anisotropic nature of $R$ hints at potential uses in applications that benefit from directional light transmission, making PolyPyGY a candidate for selective optical coatings or filters. The refractive index ($\eta$) in Figure \ref{fig:optical}(c) also displays polarization-dependent behavior, gradually decreasing with photon energy and converging near 1.0 at higher energies (around 10 eV). This trend indicates birefringent properties in PolyPyGY, once $\eta$ varies with light polarization. This anisotropic trend enhances the PolyPyGY's versatility for applications rely on controlled light propagation. 

\begin{figure*}[!htb]
    \centering
    \includegraphics[width=0.7\linewidth]{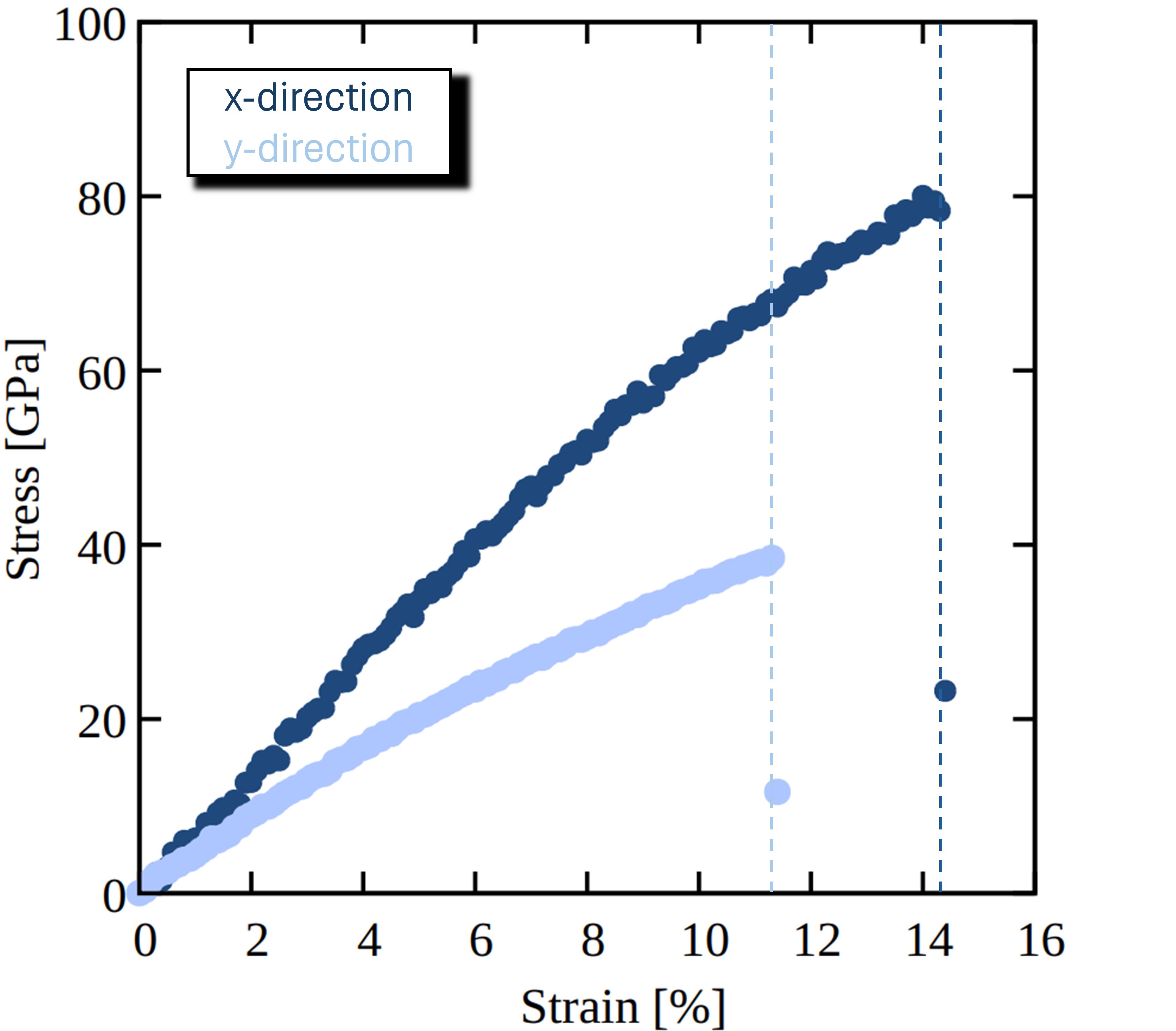}
    \caption{Stress-strain curve of PolyPyGY under uniaxial stress in the $x$-direction (dark blue) and $y$-direction (light blue), calculated using the MLIP method.}
    \label{fig:mechprop}
\end{figure*}

The mechanical response of PolyPyGY under uniaxial stress evaluated using the MLIP method is illustrated in Figure \ref{fig:mechprop}. The stress-strain curve, plotted for the $x$-direction (dark blue) and $y$-direction (light blue), reveals the material's anisotropic mechanical responses under uniaxial tensile loading, as evidenced by the distinct responses along these two orientations. The stress increases linearly with strain initially, indicating elastic behavior. At higher strain values, a peak is observed, representing the maximum stress (ultimate strength) that PolyPyGY can withstand before undergoing fracture.

In the $x$-direction, PolyPyGY reaches maximum (ultimate) stress at approximately 80 GPa for 14.3\% of strain (fracture strain, dashed dark blue line). In contrast, the ultimate stress is substantially different in the $y$-direction, reaching approximately 40 GPa at 11.3\% of strain (fracture strain, dashed light blue line). This anisotropy in the mechanical response is likely due to the structural orientation of PolyPyGY's ring configuration, which affects bond stretching and deformation under applied stress.

From the stress-strain curves shown in Figure \ref{fig:mechprop}, Young's modulus of PolyPyGY is determined to be 421.90 GPa in the $y$-direction and 663.48 GPa in the $x$-direction, also indicating anisotropic mechanical behavior. These values suggest that PolyPyGY is a relatively stiff material. However, Young's modulus is lower than that of graphene, which has a modulus close to 1 TPa in both directions \cite{scarpa2009effective,lee2012estimation}. This difference can be attributed to PolyPyGY's porous, multi-ring structure, which, while providing rigidity, introduces greater flexibility compared to graphene's continuous, denser hexagonal network. The anisotropy in PolyPyGY, with higher stiffness along the $x$-direction, reflects the alignment of its ring structures, influencing its ability to accommodate deformation.

\begin{figure*}
    \centering
    \includegraphics[width=0.8\linewidth]{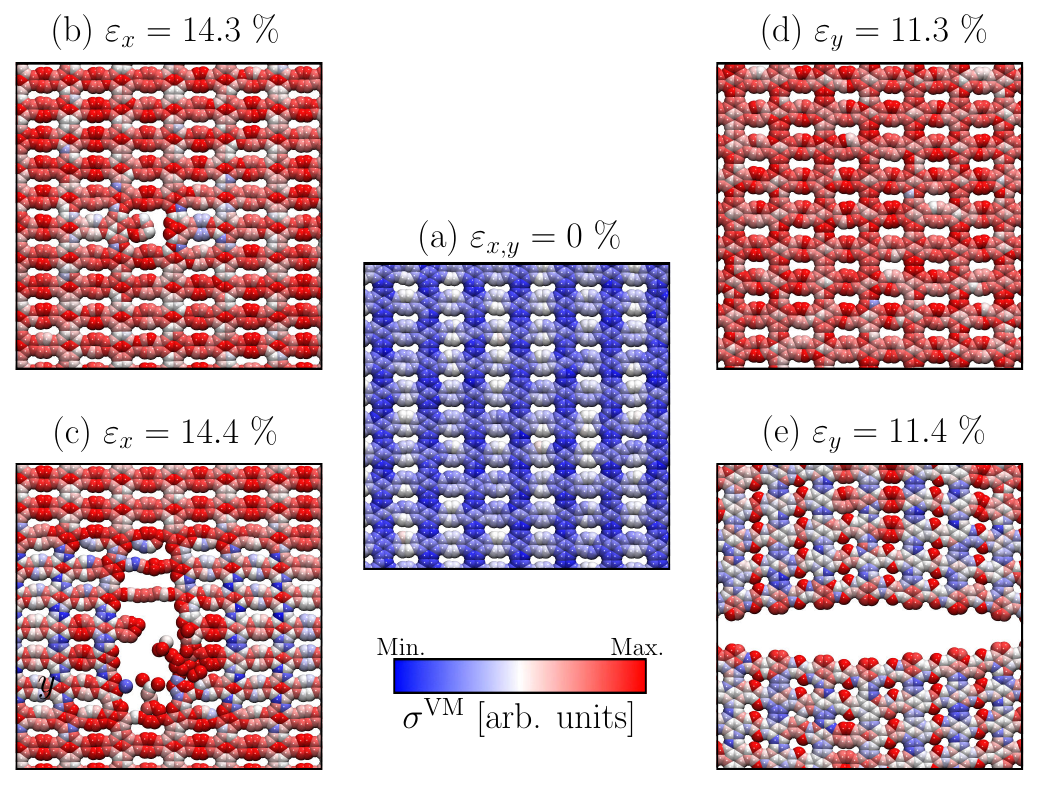}
    \caption{Fracture patterns and stress distribution in PolyPyGY, initially unstressed in panel (a), under uniaxial strain in the $x$-direction (panels b and c) and $y$-direction (panels d and e). The von Mises stress distribution is illustrated with a color gradient, where blue represents low stress and red indicates high stress.}
    \label{fig:snapshots}
\end{figure*}

Figure \ref{fig:snapshots} illustrates the fracture patterns and stress distribution in PolyPyGY under uniaxial strain applied in the x and y directions. The von Mises stress ($\sigma^\text{VM}$) distribution \cite{Lian2017Combined}, shown with a color gradient from blue (low stress) to red (high stress), reveals how strain impacts different regions of the lattice at various deformation stages, highlighting the material's fracture behavior.

The initially unstressed PolyPyGY lattice is shown in Figure \ref{fig:snapshots}(a). In Figure \ref{fig:snapshots}(b), where the structure is stretched along the $x$-direction, stress concentrations initially form around the larger rings and multi-ring junctions, and the first bonds break for a fracture of strain of 14.3\%. By 14.4\% strain (Figure \ref{fig:snapshots}(c)), these stress concentrations initiate micro-cracks in a parallel direction ($y$-direction) regarding the tensile loading direction. These micro-cracks propagate for higher strain ($\epsilon$) values, coalescing into more extensive fractures and eventually causing significant rupture across the lattice. This behavior indicates that the structural features aligned along the $x$-direction, such as the larger rings, act as weak points under tensile stress, guiding the fracture pathways along these regions.

With strain applied along the $y$-direction, a similar pattern emerges but with different fracture dynamics due to the orientation of the rings relative to the applied stress. At 11.3\% strain (see Figure \ref{fig:snapshots}(d)), stress accumulates more uniformly, but by 11.4\% strain, localized stress concentrations lead to fast crack propagation perpendicularly to the strain. These areas experience significant bond rupture, leading to an extended network of fractures. The directional dependence observed in the fracture patterns highlights the anisotropic nature of PolyPyGY, as the structure’s response to stress varies markedly between the x and y directions.

\begin{figure*}
    \centering
    \includegraphics[width=\linewidth]{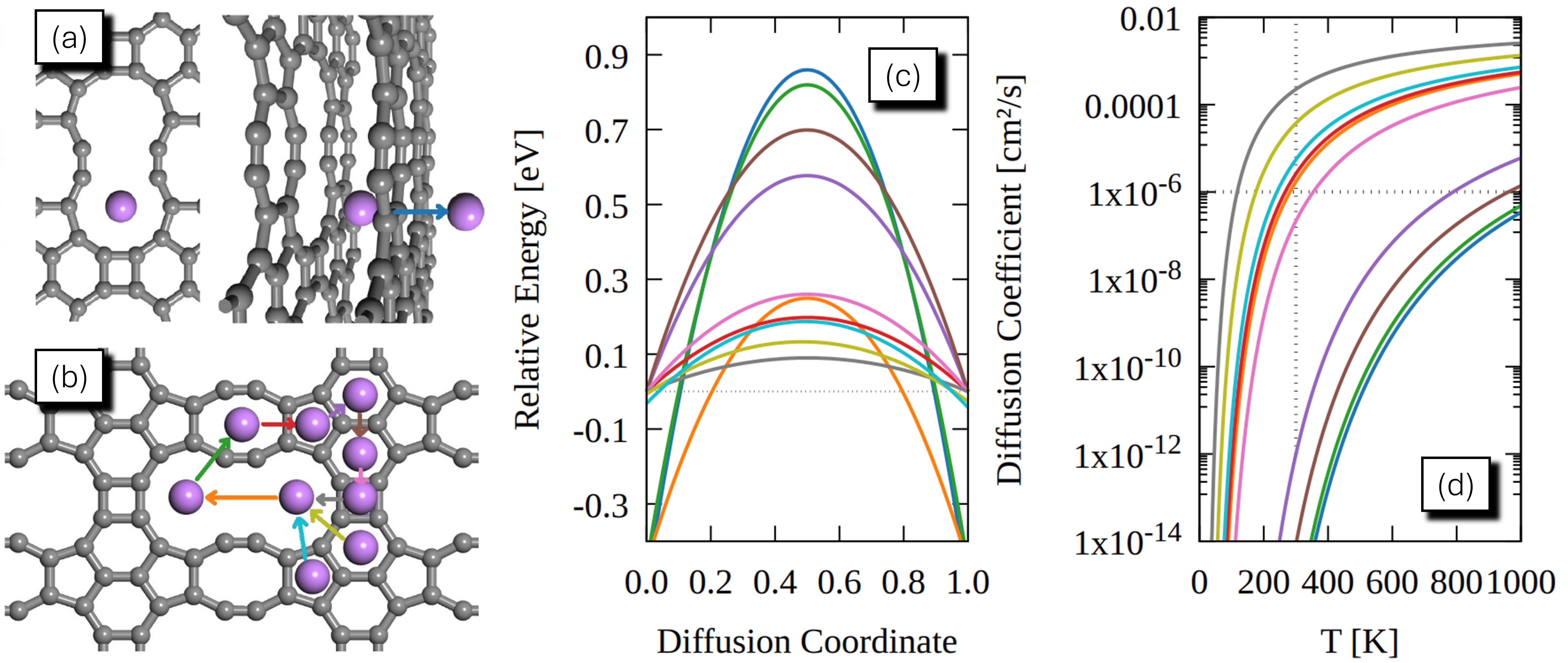}
    \caption{Lithium adsorption and diffusion properties of PolyPyGY. (a) Adsorption energy landscape and pathways for a single Li atom across and (b) on the PolyPyGY surface, (c) energy profile for Li diffusion along a selected pathway, and (d) temperature-dependent diffusion coefficients. The dashed lines in panel (d) represent the graphene diffusion coefficient at room temperature. The colored lines in panels (c) and (d) refer to the colored arrows in panels (a) and (b).}
    \label{fig:lithium}
\end{figure*}

Finally, we discuss the Li-ion adsorption properties of PolyPyGY. Figure \ref{fig:lithium} provides a comprehensive analysis of the lithium adsorption and diffusion behavior on the PolyPyGY surface. Figures \ref{fig:lithium}(a) and \ref{fig:lithium}(b) show the adsorption energy landscapes with pathways for a single Li atom diffusion across and on the PolyPyGY surface, respectively, revealing preferential adsorption sites with negative adsorption energies. The calculated adsorption energies range between $-2.3$ to $-0.93$ eV, indicating strong binding of Li to specific lattice regions. These values are comparable to \cite{zhang2024li,kim2014hydrogen,dewangan2023lithium,bi2022density} or better \cite{alvarez2021ab,wu2022potential,santos2024proposing,santos2024photh,wang2019planar,li2021two} than other 2D carbon allotropes (including graphene, about $-1.57$ eV \cite{shaidu2018lithium}), demonstrating PolyPyGY's ability to stably adsorb Li ions without compromising structural integrity. 

As shown in Figure \ref{fig:lithium}(c), energy profiles for lithium diffusion highlight the low barriers facilitating efficient ion mobility across the surface. The dashed line marks the zero point. Importantly, each pathway's minimum energy path profile was fitted using a second-order polynomial equation containing five points, two on the left and two on the right of the maximum value. The colored lines refer to the colored arrows in Figures \ref{fig:lithium}(a,b). Varying between 0.05 and 0.9 eV, these diffusion barriers are among the lowest observed for other 2D carbon allotropes \cite{zhang2024li,kim2014hydrogen,dewangan2023lithium,bi2022density,alvarez2021ab,wu2022potential,santos2024proposing,santos2024photh,wang2019planar,li2021two}, reflecting the synergy between PolyPyGY's porous network and its electronic structure. The barriers' directional dependence emphasizes the material's anisotropic nature, which could be leveraged to optimize ion transport along specific pathways. 

Figure \ref{fig:lithium}(d) depicts temperature-dependent diffusion coefficients \cite{gomez2024tpdh} that underscore PolyPyGY's performance under various operating conditions. The dashed lines represent the graphene diffusion coefficient at room temperature for comparison. The colored lines refer to the colored arrows in Figures \ref{fig:lithium}(a,b). At room temperature, the diffusion coefficient surpasses $6\times 10^{-6}$ cm$^{2}$/s, indicative of excellent Li-ion mobility \cite{kucinskis2013graphene,wang2009graphene}. As temperature increases, the exponential growth in diffusion coefficients confirms the thermal activation of Li-ion dynamics, further validating PolyPyGY's suitability for high-efficiency LIBs.

\begin{figure*}
    \centering
    \includegraphics[width=0.8\linewidth]{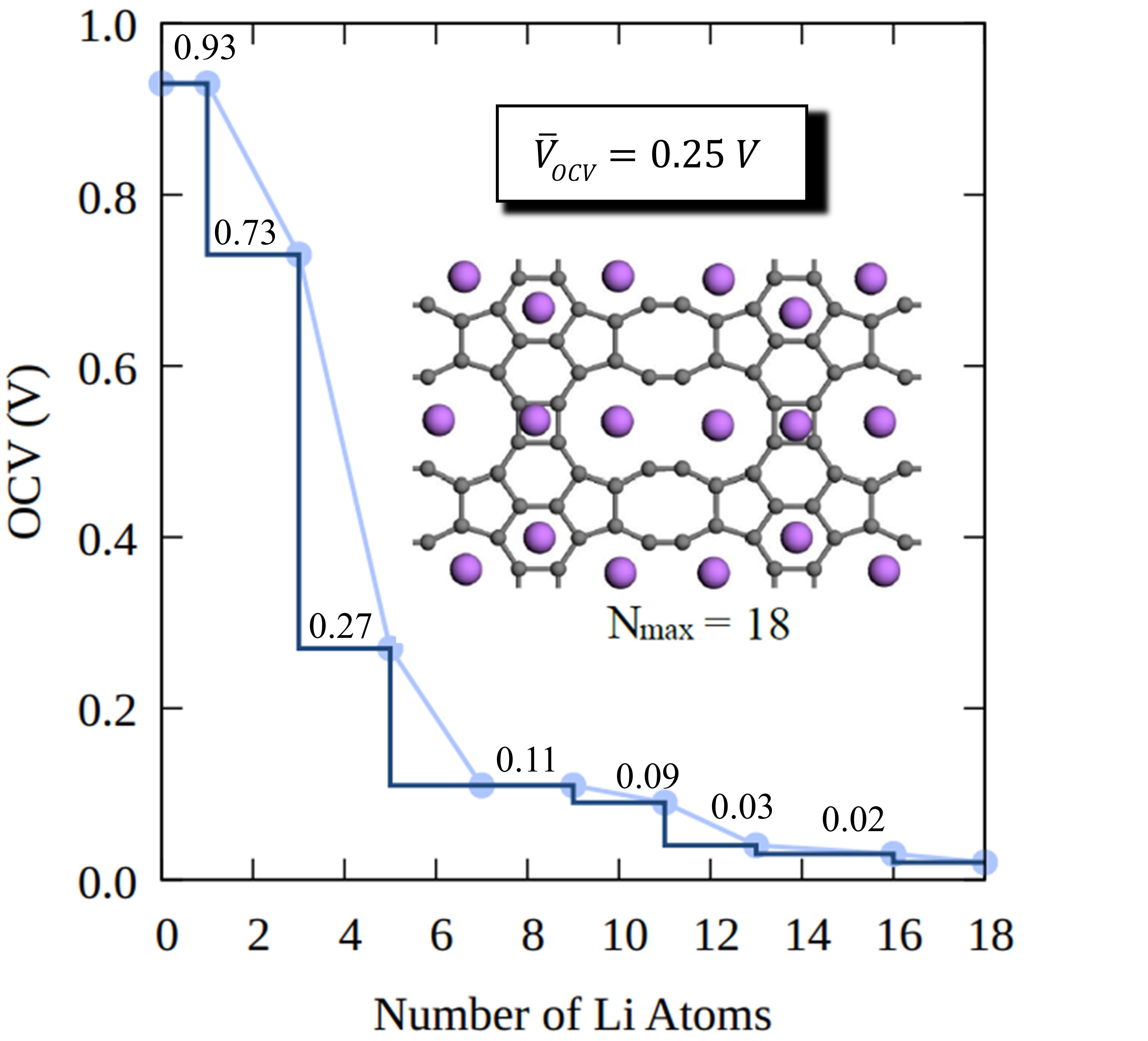}
    \caption{OCV as functions of the number of adsorbed in PolyPyGY. The average OCV is 0.25 eV.}
    \label{fig:ocv}
\end{figure*}

The interplay between open circuit voltage (OCV) and the number of lithium atoms is presented in Figure \ref{fig:ocv}. The OCV values were calculated using the method described in reference \cite{gomez2024tpdh}. One can note a clear trend as adsorbed lithium atoms increase. The initial adsorption energy is highly negative, indicating strong binding at low Li coverage. As the lithium coverage increases, the adsorption energy gradually increases, stabilizing near $-2.3$ eV at 18 Li atoms, as mentioned above. This behavior reflects the increasing electrostatic repulsion and site competition among Li atoms as more are adsorbed. Despite this, PolyPyGY maintains strong adsorption even at high lithium concentrations, suggesting that its porous and multi-ringed structure effectively accommodates Li ions without compromising binding stability.

PolyPyGY's average OCV is approximately 0.25 eV, positioning it among materials with relatively low OCV values compared to other 2D carbon-based allotropes. This trend makes it attractive for acting as an anode in Li-ion battery applications. Low OCVs are generally better for anodes since they minimize the voltage difference between the anode and the Li-ion agglomerate, improving efficiency. A low OCV also reduces the risk of lithium plating during charging and ensures the battery can store more energy effectively.

Graphene, for instance, exhibits an OCV of 0.11 eV \cite{liu2012folded}, while graphite has an OCV ranging from 0.22 to 0.40 eV \cite{persson2010thermodynamic}. These values suggest that PolyPyGY demonstrates comparable electrochemical stability and is suitable for Li-ion battery applications. However, compared to TPDH-graphene and twin-graphene, which exhibit OCVs as low as 0.29 eV \cite{gomez2024tpdh}
and 0.32 eV \cite{gao2023twin}, respectively, PolyPyGY offers a competitive balance between stability and energy density. Materials with higher OCV values, such as QPHT-graphene (0.66 eV) \cite{qiu2022qpht} and popgraphene (0.45 eV) \cite{wang2018popgraphene}, are more suitable for cathode applications but may present challenges for anode usage due to their higher initial voltage requirements. The low average OCV of PolyPyGY reflects its potential as an anode material, ensuring a stable intercalation voltage that is less likely to lead to lithium plating during charge cycles. This property, combined with its metallic nature and favorable diffusion barriers, underscores PolyPyGY's versatility and efficacy in advanced energy storage systems.

The theoretical capacity of PolyPyGY in storing lithium ions, calculated to be 2231.41 mAh/g, far surpasses some of the 2D carbon-based allotropes referenced here. For example, graphene exhibits a much lower capacity of 568 mAh/g \cite{liu2012folded}, and graphite, commonly used in commercial lithium-ion batteries, only has 372 mAh/g capacity \cite{persson2010thermodynamic}. Among the moderate-capacity materials, graphenylene and popgraphene offer capacities of 1116 mAh/g \cite{yu2013graphenylene} and 1487 mAh/g \cite{wang2018popgraphene}, respectively. However, these values still fall significantly short of PolyPyGY's capacity. Twin-graphene, for instance, exhibits a remarkable capacity of 3916 mAh/g \cite{gao2023twin}, much higher than PolyPyGY.

PolyPyGY's good capacity can be attributed to its unique multi-ringed porous structure, which provides abundant adsorption sites for lithium ions. This structural advantage makes PolyPyGY a competitive candidate for next-generation lithium-ion battery anodes, combining high theoretical capacity with its moderate OCV (0.25 eV). Such a balance between capacity and electrochemical stability is crucial for ensuring high energy density and long-term cycling performance in practical applications.

\section{Conclusions}

This study presented Polymerized Pyracyclene Graphyne, a novel 2D carbon allotrope, as a promising candidate for LIB anodes due to its unique structural, electronic, mechanical, and electrochemical properties. PolyPyGY's multi-ringed architecture, composed of 4-, 5-, 6-, 8-, and 16-membered carbon rings, ensures structural stability and provides high porosity and diverse adsorption sites, critical for efficient lithium storage and transport. We demonstrated PolyPyGY's excellent structural stability through DFT and machine learning approaches. The material exhibits anisotropic mechanical properties. Young's modulus values of 421 GPa in the $y$-direction and 664 GPa in the $x$-direction offer both rigidity and flexibility suitable for diverse applications.

PolyPyGY's electronic structure, characterized by metallic behavior, ensures efficient charge transport, while the strong adsorption energies ($-2.3$ to $-0.93$ eV) and low diffusion barriers (0.05 to 0.9 eV) facilitate rapid Li-ion mobility. The material's high diffusion coefficient, exceeding $6\times 10^{-6}$ cm$^{2}$/s at 300 K, underscores its ability to support fast charge and discharge cycles. Furthermore, its average open circuit voltage (0.25 eV), stabilizing around 0.02 eV at higher lithium coverage, and its theoretical capacity for Li-ion storing (about 2231.41 mAh/g) confirm its suitability for high-capacity energy storage applications.

\section{Acknowledgements}
\noindent This work received partial support from Brazilian agencies CAPES, CNPq, and FAPDF.
L.A.R.J. acknowledges the financial support from FAP-DF grants 00193.00001808/2022-71 and $00193-00001857/2023-95$, FAPDF-PRONEM grant 00193.00001247/2021-20, PDPG-FAPDF-CAPES Centro-Oeste 00193-00000867/2024-94, and CNPq grants $350176/2022-1$ and $167745/2023-9$. 
M.L.P.J. acknowledges the financial support of the FAP-DF grant 00193-00001807/2023-16. M.L.P.J. and L.A.R.J. thanks also to CENAPAD-SP (National High-Performance Center in São Paulo, State University of Campinas -- UNICAMP, projects: proj960 and proj634, respectively) and NACAD (High-Performance Computing Center, Lobo Carneiro Supercomputer, Federal University of Rio de Janeiro -- UFRJ, projects: a22002 and a22003, respectively) for the computational support provided. The authors acknowledge the National Laboratory for Scientific Computing (LNCC/MCTI, Brazil) for providing HPC resources for the SDumont supercomputer, contributing to the research results reported in this paper.

\section{Supporting Information}

\bibliographystyle{unsrt}
\bibliography{references}
\end{document}